\documentclass[manuscript]{acmart}
\AtBeginDocument{%
  }

\usepackage{enumitem}
\usepackage{graphicx}
\usepackage{subcaption}
\usepackage{caption}

\setcopyright{acmlicensed}
\copyrightyear{2026}
\acmYear{2026}
\setcopyright{cc}
\setcctype{by}
\acmConference[IH '26]{Interactive Health Conference}{July 05--08, 2026}{Porto, Portugal}
\acmBooktitle{Interactive Health Conference (IH '26), July 05--08, 2026, Porto, Portugal}
\acmDOI{10.1145/3786579.3804960}
\acmISBN{979-8-4007-2422-0/2026/07}

\begin{document}

\title{Exploring Expert Perspectives on Wearable-Triggered LLM Conversational Support for Daily Stress Management}

\author{Poorvesh Dongre}
\affiliation{%
  \institution{Harvard Medical School}
  \city{Boston}
  \country{USA}
}
\email{pdongre1@mclean.harvard.edu}

\author{Sameer Neupane}
\affiliation{%
  \institution{University of Memphis}
  \city{Memphis}
  \country{USA}}
 \email{sameer.neupane@memphis.edu}

\author{Priyanka Jadhav}
\affiliation{%
  \institution{Omnissa}
  \city{Foster City}
  \country{USA}}
 \email{priyankaj72@gmail.com}

\author{Nikitha Donekal Chandrashekar}
\affiliation{%
  \institution{Virginia Tech}
  \city{Blacksburg}
  \country{USA}}
 \email{nikitha@vt.edu}

\author{Christian Webb}
\affiliation{%
  \institution{Harvard Medical School}
  \city{Boston}
  \country{USA}}
 \email{cwebb@mclean.harvard.edu}

\author{Santosh Kumar}
\affiliation{%
  \institution{University of Memphis}
  \city{Memphis}
  \country{USA}}
 \email{santosh.kumar@memphis.edu}

\author{Denis Gra{\v{c}}anin}
\affiliation{%
  \institution{Virginia Tech}
  \city{Blacksburg}
  \country{USA}}
 \email{gracanin@vt.edu}

\renewcommand{\shortauthors}{Dongre et al.}

\begin{abstract}
Wearable devices increasingly support stress detection, while LLMs enable conversational mental health support.
However, designing systems that meaningfully connect wearable-triggered stress events with generative dialogue remains underexplored, particularly from a design perspective.
We present EmBot, a functional mobile application that combines wearable-triggered stress detection with LLM-based conversational support for daily stress management.
We used EmBot as a design probe in semi-structured interviews with 15 mental health experts to examine their perspectives and surface early design tensions and considerations that arise from wearable-triggered conversational support, informing the future design of systems for daily stress management and mental health support.
\end{abstract}




\maketitle
 
\section{Introduction \& Background}

Stress is a mental state arising from cognitive or emotional overload when demands exceed an individual's capacity. 
While short-term stress can be helpful, chronic stress is associated with adverse mental health outcomes such as mood and eating disorders and depression~\cite{calabrese2009neuronal,troop1998stress,hammen2005stress}, and negatively affects well-being and productivity~\cite{dougall2001stress,erickson2009severity}. 

Over the past decade, technological advances have enabled systems for daily stress management. 
Passive health sensing applications, particularly wearables such as wristbands, smartwatches, and rings, capture physiological and behavioral signals, which are used to detect and monitor stress~\cite{ollander-2016-stress-empatica,gjoreski2016continuous,wang2018tracking,wong-2019-stress-empatica,shah2021personalized,chandra2021comparative,abd2023wearable,Kuzmowycz_2023}. 
Beyond passive sensing, wearables can also provide stress interventions, including biofeedback~\cite{neupane2024momentary,sharmin2015visualization,sanches2010mind,kocielnik2013smart} and just-in-time adaptive prompting~\cite{sensor-stress-management2015,sarker2016finding, smith2020integrating,battalio2021sense2stop}. 
However, prior work highlights challenges including false positives, notification fatigue, disengagement, and contextual ambiguity in wearable data~\cite{neupane2025wearable, neupane2024momentary}. 
These challenges suggest that stress detection and monitoring alone is insufficient; a key unresolved challenge is how to translate wearable data into timely, meaningful, and user-appropriate interactions. 

LLMs introduce new opportunities for daily stress monitoring and intervention through their natural language inference and generative capabilities. 
Prior work explores LLMs for emotion logging~\cite{singh2025annosense}, therapy chatbots~\cite{heinz2025randomized}, and summarizing wearable-derived insights for mental health support~\cite{10.1145/3749474}. 
LLMs have also been used for conversational mental health support, including therapeutic-style dialogue~\cite{liu2023chatcounselor, lai2023psy} and generative journaling or coaching systems~\cite{nepal2024mindscape,wang2025exploring}. 
However, many LLM-based mental health systems operate independently of wearable sensing, relying primarily on user-provided text. 
As a result, they lack grounding in wearable data, limiting their ability to proactively and empathically support users in everyday contexts. 

Recent efforts have investigated using LLMs for sensemaking of wearable data for activity~\cite{fang2024physiollm}, sleep~\cite{10.1145/3706598.3713852}, and general health \& behavioral health~\cite{10.1145/3749474}. 
These systems primarily focus on post-hoc interpretation or summarization of sensed data rather than supporting real-time, user-facing interaction. 
In parallel, emerging systems have explored wearable-triggered conversational support for stress management~\cite{neupane2025wearable,dongre2024physiology,dongre2025empathic}. 
While promising, these efforts largely demonstrate feasibility and do not yet provide a clear understanding of how such interactions should be designed in real-world mental health contexts. 
In particular, the role of mental health experts in shaping these interactions remains underexplored which is essential in ensuring safety, appropriateness, and clinical relevance in early-stage system design. 

To address this gap, we developed EmBot (short for Empathic Chatbot), a functioning mobile application that uses wearable-triggered stress events to initiate and ground LLM-based conversational support and used it as a design probe to ground discussions with 15 mental health experts (researchers and clinicians) on the design of wearable-triggered LLM conversational systems for daily stress management. 
Experts interacted with EmBot (either through guided hands-on use or a structured walkthrough) and evaluated its interaction flow, triggering mechanisms, feedback design, and potential real-world applicability. 
Our goal was to elicit expert perspectives to inform early-stage design decisions for wearable-triggered LLM conversational systems, with a focus on how such systems should be designed, triggered, and integrated into real-world mental health contexts.

\section{System Design}

\begin{figure*}[t]
    \centering
    \begin{subfigure}[t]{0.175\textwidth}
        \includegraphics[width=\linewidth]{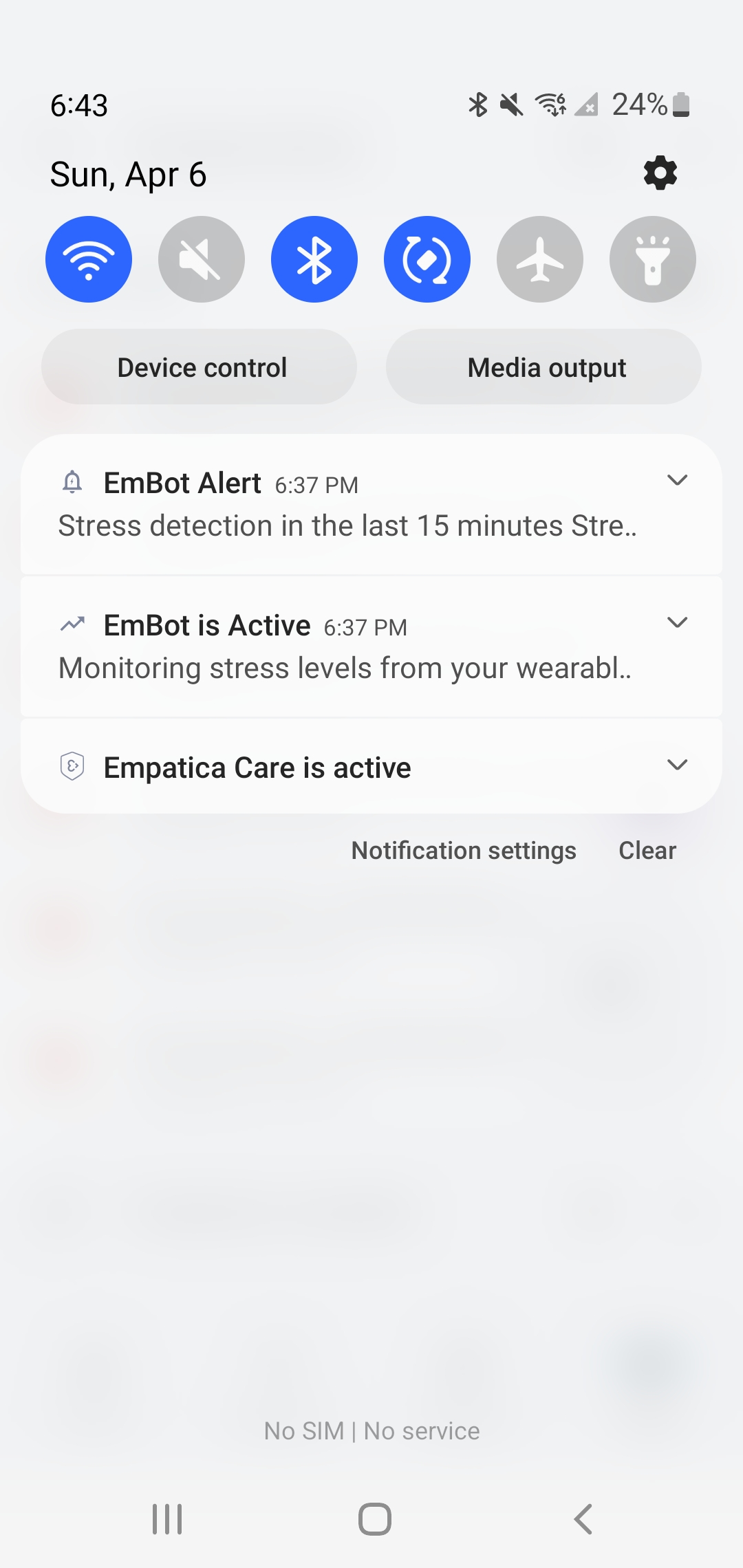}
        \caption{Detection: Notification when stress is detected.}
        \label{fig:notification}
    \end{subfigure}
    \hfill
    \begin{subfigure}[t]{0.175\textwidth}
        \includegraphics[width=\linewidth]{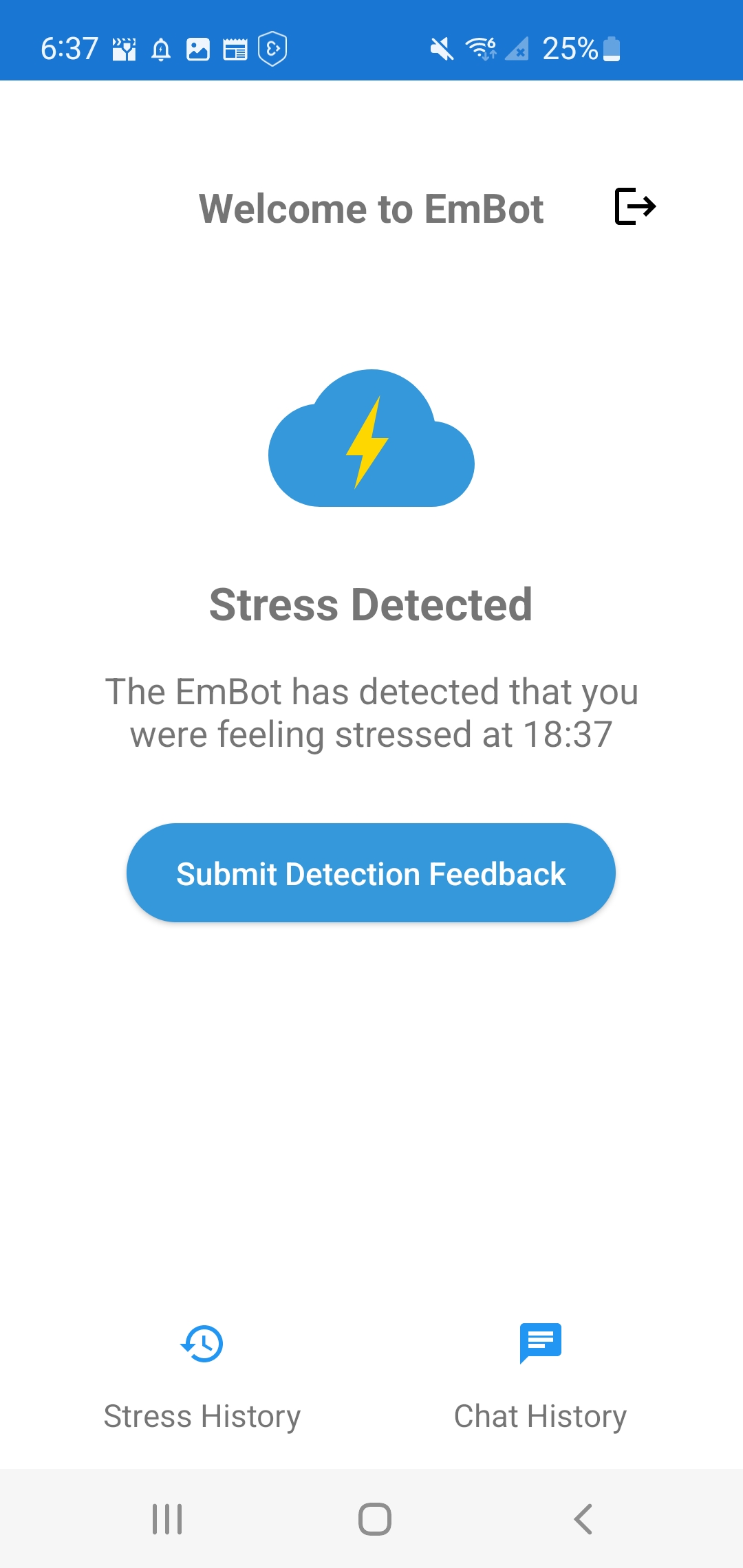}
        \caption{Detection: Screen with real-time stress detection.}
        \label{fig:home}
    \end{subfigure}
    \hfill
    \begin{subfigure}[t]{0.175\textwidth}
        \includegraphics[width=\linewidth]{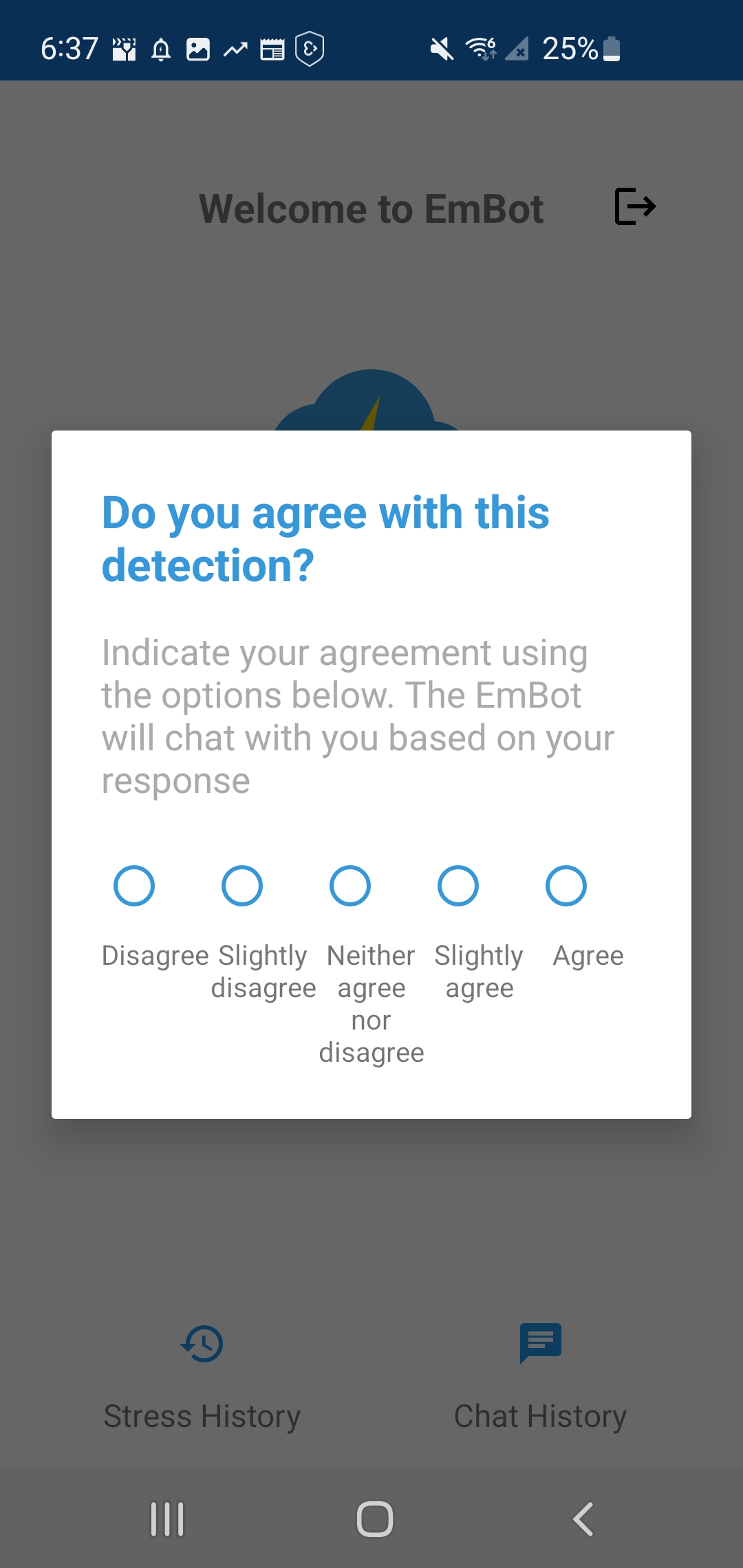}
        \caption{Feedback: User stress detection agreement scale.}
        \label{fig:feedback}
    \end{subfigure}
    \hfill
    \begin{subfigure}[t]{0.175\textwidth}
        \includegraphics[width=\linewidth]{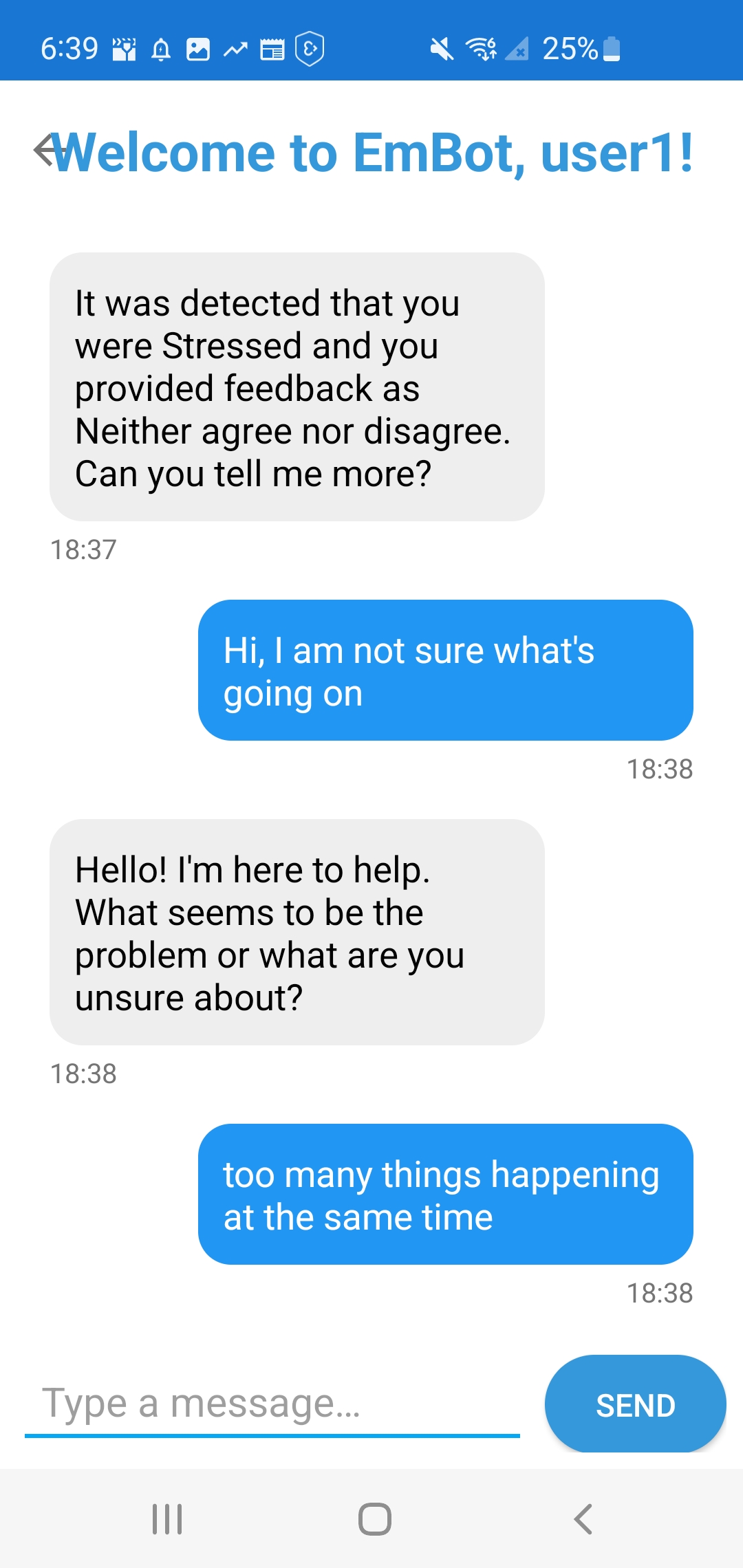}
        \caption{Support: In-app chat support powered by LLM.}
        \label{fig:chat_history}
    \end{subfigure}
    \hfill
    \begin{subfigure}[t]{0.175\textwidth}
        \includegraphics[width=\linewidth]{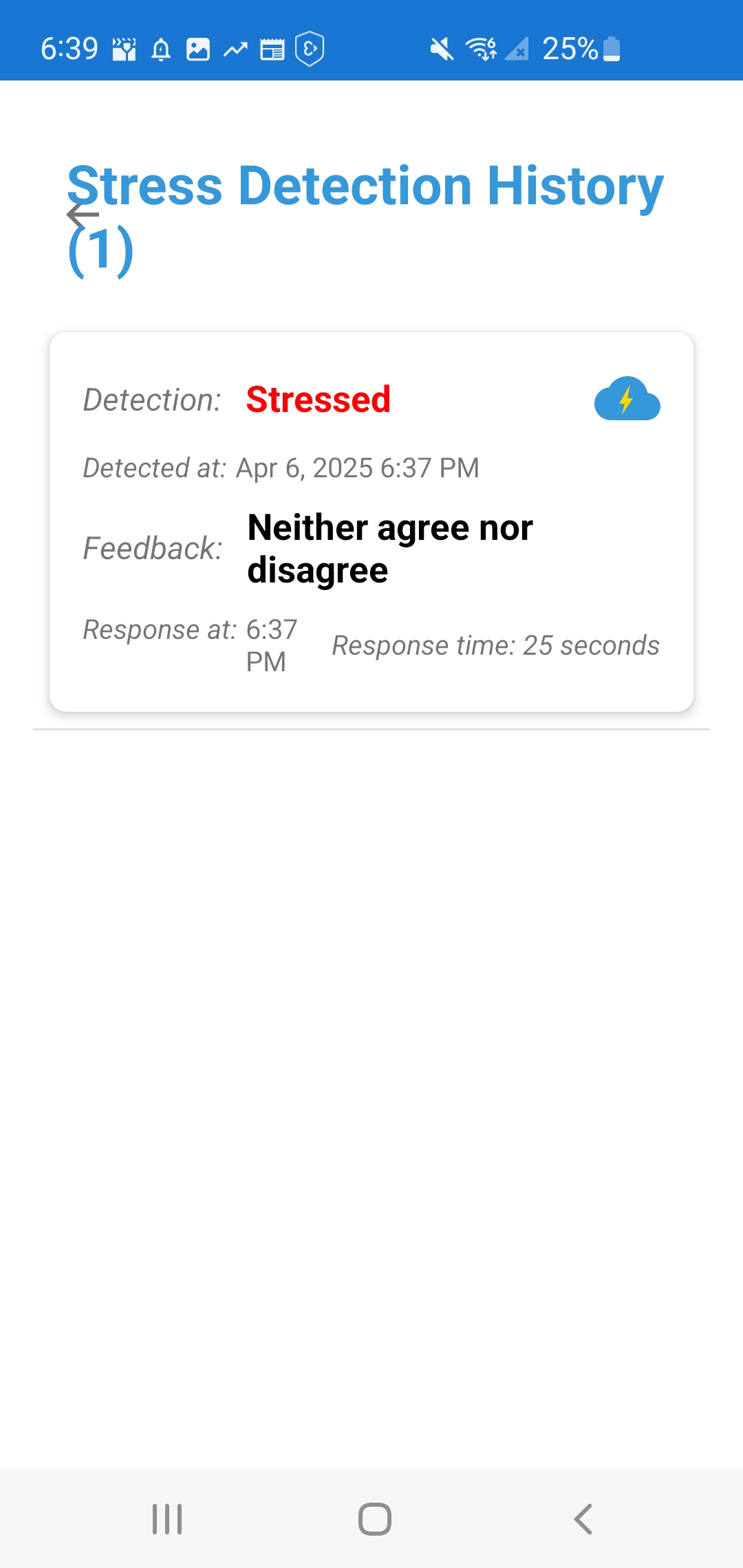}
        \caption{Reflection: History with all stress detections.}
        \label{fig:stress_history}
    \end{subfigure}
    \caption{Interaction Stages in EmBot: Detection, Feedback, Support, and Reflection.}
     \Description{Interaction stages when using EmBot. Detection stage includes a notification when stress is detected followed by real-time stress detection. It is followed by Feedback stage complemented by an LLM-based in-app chat (Support stage). Finally, Reflection stage provides a history of all stress detections.}
    \label{fig:embot}
\end{figure*}

EmBot (\autoref{fig:embot}) was developed as a functional mobile design probe to support grounded discussion during expert interviews. 
The system was implemented to allow experts to experience a plausible interaction flow of a wearable-triggered LLM conversational system for daily stress management. 
As a design probe, EmBot was not intended to evaluate model performance, but to elicit interaction-level feedback grounded in a concrete user experience. 

The mobile application is paired with a wearable device that monitors raw physiological signals and detects stress (\autoref{fig:notification}). 
For the interviews, stress events were simulated to ensure consistent scenarios across participants and to focus discussion on interaction design rather than model accuracy. 
When a notification appears, users can review and respond to the detected stress event (\autoref{fig:home} and \autoref{fig:feedback}). 
This interaction stage was designed to preserve user agency by allowing users to confirm, reject, or contextualize the detection before proceeding. 

Following user feedback, EmBot initiates a conversational interaction powered by an LLM (\autoref{fig:chat_history}). 
The conversation is grounded in the detected event and user input, offering reflective prompts and coping-oriented dialogue. 
Rather than presenting static advice, the probe demonstrates how wearable data can be translated into conversational engagement.
To support longer-term reflection, users can revisit a history of stress detections and prior conversations (\autoref{fig:stress_history}). 
This feature allows exploration of how wearable-triggered LLM systems might support stress management and sustained engagement over time. 

\section{Methodology}

We conducted semi-structured interviews with 15 mental health experts (18 interviews conducted; 3 excluded due to recording issues). 
These included licensed clinicians, clinical researchers, and computer scientists whose primary research focus is digital mental health. 
Experts were initially recruited via professional networks and direct outreach to researchers and clinicians in digital mental health, followed by snowball sampling through participant referrals. 

As discussed, EmBot was used as a functional design probe during interviews. 
All experts were exposed to the same four interaction stages: Detection, Feedback, Support, and Reflection. 
As discussed, stress detection events and associated data were simulated and experts were instructed to interpret these scenarios as plausible real-world situations, allowing them to evaluate system behavior and design implications independent of model accuracy. 

Interviews were conducted in-person and remotely, with both groups first viewing a structured walkthrough video that demonstrated each interaction stage. 
Following this, in-person experts interacted with the fully functional mobile application, while remote experts used an interactive mockup that replicated the same interaction flows. 
In both formats, experts were encouraged to ask questions, request alternative scenarios, and discuss hypothetical use cases. 

Each interview lasted 45--60 mins and was conducted under IRB approval with informed consent. 
Each session followed a semi-structured format consisting of two major phases: pre- and post-probe. 
The pre-probe phase included background discussion on experts' experience with wearables and LLMs for mental health (approx 5-10 mins) and experts' view of wearable-triggered LLM conversational systems for daily stress management (10--15 mins). 
The post-probe phase involved a guided walkthrough and interaction with EmBot (approx 15--20 mins), followed by a reflective discussion on design implications, risks, safety considerations, personalization, and clinical appropriateness (approx 15--20 mins). 

Interviews were audio-recorded, transcribed, and analyzed using reflexive thematic analysis following Braun and Clarke~\cite{braun2006using}. 
Two researchers independently conducted first-cycle open coding across transcripts to identify patterns related to system perception and interaction design. 
Coding discrepancies were discussed and resolved through consensus, with iterative refinement of the codebook throughout the analysis process. 
Through iterative comparison and discussion, overlapping codes were consolidated into higher-level categories. 
The researchers then collaboratively refined these categories into themes that captured recurrent design tensions and considerations.

\section{Findings}

\subsection{Pre-probe Perspectives}

Before interacting with EmBot, experts reflected on the well-established advantages and limitations of using wearables and LLMs for daily stress management and other mental health applications. 
However, when asked about their views on using the two together for daily stress management, experts mostly reiterated their thoughts on wearables and LLMs. 
This illustrates that, prior to engaging with EmBot, experts tended to describe wearables and LLMs as separate tools, often reiterating their standalone thoughts rather than articulating how they might work together in a unified system.

\subsection{Probe-informed Opportunities}

After interacting with EmBot's four stages (Detection, Feedback, Support, Reflection), experts began discussing more concrete interaction-level considerations on using wearables and LLMs together. 
The probe facilitated discussion on various topics, including timing, interpretability, conversational tone, and safety mechanisms. 
These considerations emerged from experts' engagement with EmBot's interaction stages. 

\subsubsection{From Passive Monitoring to Contextual Dialogue:}

When exploring EmBot's stress-triggered notification and LLM conversation, several experts emphasized the potential to transform real-time wearable detection into empathic dialogue. 
As E6 noted, ``I like that the chatbot is reaching out to you because it thinks you're stressed.'' 
Several experts described this as shifting monitoring from passive data capture to empathic engagement, creating ``opportunities to capture more ecologically valid data\ldots'' (E14). 
However, they emphasized that wearable detection and the wearable-triggered LLM conversations could be made more accurate and meaningful by grounding them in other contextual details such as, location, activity, and sleep. 

\subsubsection{Detection Transparency and Notification Calibration:}

Interaction with the notification and feedback screens prompted suggestions around interpretability and pacing. 
Several experts recommended explicitly explaining detection rationale in the triggered LLM conversations: ``Maybe the chat could say- I picked up a bit of stress, maybe your heart rate went up'' (E1). 
Some experts suggested allowing users to query the detection source in the conversation: ``Was it heart rate? Was it more movement?'' (E17). 
Notification fatigue emerged as a concrete design concern, and experts suggested capped alerts, adaptive pacing, and disengagement detection within the wearable-triggered conversations (E8). 
Because notifications in EmBot are ideally designed to be triggered by wearable detection rather than user-initiated actions, experts emphasized on transparency for maintaining user trust in both sensing accuracy and conversational intent. 

\subsubsection{Structured and Adaptive Conversational Support:}

After engaging with EmBot's LLM-driven conversations, some experts encouraged more structured follow-up questioning rather than long, generic responses.
E14 suggested asking: ``Did anything happen? Did you argue with someone?\ldots And then we move on to the baby questions.'' 
Personalization of tone based on the wearable detection was also emphasized to make the wearable-triggered conversation more empathic. 
Other suggestions include making the conversations more human chat-like by using typing indicators and using voice interactions (E2). 
These reflections are applicable to other LLM conversations as well but become particularly important in EmBot because conversations are initiated by wearable detection rather than explicit user input. 
As a result, the system must quickly establish relevance, context, and trust without prior user framing. 

\subsubsection{Supporting Reflection Without Increasing Burden:}

When reviewing EmBot's stress history view, some experts highlighted using LLMs to support the identification of stress patterns (E10). 
Importantly, they positioned wearable-triggered conversations as intermediaries between users and clinicians. 
There were concerns about wearable data increasing user and clinician burden but LLMs were seen as capable of distilling it into concise, human-readable summaries for both users and clinicians, thereby facilitating user-clinician interaction: ``It tells me\ldots you talked with your therapist\ldots do you follow that up?'' (E1). 
EmBot provides users and clinicians unique opportunities to collaborate in everyday mental health context making it different from existing wearable sense-making applications. 

\subsubsection{Transparency, Safety, and Privacy:}

Interaction with EmBot also surfaced specific transparency and safety considerations because wearable detection and generative dialogue jointly shape user interpretation, amplifying the consequences of false detections or inappropriate response. 
Several experts emphasized onboarding clarity about capabilities and limits: ``Here's what it can do\ldots and what it cannot do'' (E18). 
Experts also emphasized having user privacy control, including deletion of the detected stress events and sensitive conversations (E4). 
For high-risk scenarios, detected either with the wearable data or conversation, escalation mechanisms such as crisis resource links were strongly recommended: ``If there's a way\ldots to detect high-risk events so it can divert people'' (E18).

\section{Discussion}

Our findings surface recurring design tensions and early design considerations that emerged when experts engaged with EmBot.

\subsection{Cross-Cutting Design Tensions}

Across interviews, experts' reactions to EmBot revealed several recurring design tensions. 
Experts noted that while continuous sensing enables contextually grounded insights, it also raises concerns about intrusiveness when translated into frequent conversational prompts. 
While wearable-triggered outreach was described as supportive and empathic, experts noted that poorly calibrated triggers may affect how conversational support is perceived and continued. 
While wearable-triggered conversations opportunities for meaningful reflection, overly definitive interpretations may misrepresent ambiguous wearable data, while overly cautious responses may reduce usefulness. 
Experts also cautioned when conversations are initiated based on wearable detection, there is a risk of overeliance and users may attribute greater authority to system responses and perceived them as offering clinical or therapeutic guidance compared to generic chatbots.

\subsection{Preliminary Design Considerations}

Expert discussions suggested several preliminary considerations along which the design of wearable-triggered LLM conversational systems may vary. 
First, wearables may function primarily as passive monitoring or as an active trigger that initiates LLM conversations, directly influencing when and why interactions occur. 
Second, triggers can be wearable-driven or time-based, shaping how LLM conversations are initiated and enabling the collection of free-text ecological momentary assessments to support reflection and system adaptation. 
Third, LLMs may act as sense-makers that summarizes insights from wearable data in natural language, conversational agents that support reflective dialogue, or mediators that act as intermediaries between the users and clinicians. 
User interactions with LLMs may be wearable-triggered, user-initiated, or a combination of both, allowing user the autonomy to shape their conversation with the wearable-triggered LLM. 
Finally, the overall system may differ in intended scope, ranging from supporting everyday self-reflection to augmenting clinically relevant interactions.

\subsection{Scope, Limitations, and Future Work}

Our study represents an early-stage design exploration of wearable-driven LLM conversations focusing on daily stress management. 
The design tensions and preliminary design considerations articulated in this work do not constitute a formal framework or validated design space. 
Rather, they represent recurring trade-offs surfaced through expert engagement with a design probe. 
These insights primarily serve to refine EmBot's design before broader deployment. 

We also acknowledge the methodological limitations of our study. 
First, experts engaged with EmBot in person or remotely, and although all experts experienced the same interaction stages, differences in modality may have influenced the depth or immediacy of feedback. 
Second, stress detection within EmBot was treated as operational and, for demonstration purposes, simulated rather than empirically validated. 
Third, our sample included experts with diverse backgrounds, which enriched perspectives but may also reflect varying assumptions about clinical deployment and technical feasibility. 

Future work will focus on refining and deploying EmBot with end users. 
This includes examining real-world stress detection, engagement, and usefulness (clinical and perceived); evaluating notification calibration strategies over extended periods; assessing conversational adaptation and personalization in practice; and studying how hybrid systems integrate into existing clinical workflows without increasing burden.

\section{Conclusion}

This work presents EmBot as a functional design probe to explore how wearable-triggered LLM conversational support can be meaningfully designed and deployed for daily stress management and other mental health applications. 
Through interviews with mental health experts, we identified how engagement with EmBot as a functional design probe surfaces preliminary design tensions and considerations that extend beyond abstract discussions of wearables or LLM chatbots in isolation. 
Our contribution is exploratory, demonstrating how probe-based engagement can reveal key challenges and opportunities inherent to combining wearable sensing with generative dialogue. 
These insights inform the iterative refinement of EmBot and provide early design guidance for future longitudinal deployments of wearable-triggered LLM conversational systems in daily stress management.

\section{Disclosures}

We used ChatGPT (GPT-5.2) to assist in the writing of this manuscript, including fixing grammar and refining sentences. 

\bibliographystyle{ACM-Reference-Format}
\bibliography{sample_bibliography}

\end{document}


\title{Exploring Expert Perspectives on Wearable-Triggered LLM Conversational Support for Daily Stress Management}

\author{Poorvesh Dongre}
\affiliation{%
  \institution{Harvard Medical School}
  \city{Boston}
  \country{USA}
}
\email{pdongre1@mclean.harvard.edu}

\author{Sameer Neupane}
\affiliation{%
  \institution{University of Memphis}
  \city{Memphis}
  \country{USA}}
 \email{sameer.neupane@memphis.edu}

\author{Priyanka Jadhav}
\affiliation{%
  \institution{Omnissa}
  \city{Foster City}
  \country{USA}}
 \email{priyankaj72@gmail.com}

\author{Nikitha Donekal Chandrashekar}
\affiliation{%
  \institution{Virginia Tech}
  \city{Blacksburg}
  \country{USA}}
 \email{nikitha@vt.edu}

\author{Christian Webb}
\affiliation{%
  \institution{Harvard Medical School}
  \city{Boston}
  \country{USA}}
 \email{cwebb@mclean.harvard.edu}

\author{Santosh Kumar}
\affiliation{%
  \institution{University of Memphis}
  \city{Memphis}
  \country{USA}}
 \email{santosh.kumar@memphis.edu}

\author{Denis Gra{\v{c}}anin}
\affiliation{%
  \institution{Virginia Tech}
  \city{Blacksburg}
  \country{USA}}
 \email{gracanin@vt.edu}

\renewcommand{\shortauthors}{Dongre et al.}

\begin{abstract}
Wearable devices increasingly support stress detection, while LLMs enable conversational mental health support.
However, designing systems that meaningfully connect wearable-triggered stress events with generative dialogue remains underexplored, particularly from a design perspective.
We present EmBot, a functional mobile application that combines wearable-triggered stress detection with LLM-based conversational support for daily stress management.
We used EmBot as a design probe in semi-structured interviews with 15 mental health experts to examine their perspectives and surface early design tensions and considerations that arise from wearable-triggered conversational support, informing the design of future wearable-triggered LLM conversational systems for daily stress management.
\end{abstract}

\ccsdesc[500]{Human-centered computing~Empirical studies in ubiquitous and mobile computing}
\ccsdesc[500]{Human-centered computing~Ubiquitous computing}
\ccsdesc[500]{Human-centered computing~User studies}


\keywords{mental health, wearables, stress, generative AI, large language models, chatbot}



\begin{table}[t]
\small
\centering
\caption{Expert role and expertise.}
\label{tab:participants}
\begin{tabular}{|c|p{4.2cm}|p{10cm}|}
\hline
\textbf{ID} & \textbf{Role} & \textbf{Area of Expertise} \\
\hline
E1 & Clinical Psychologist & Cognitive behavior therapy, lifestyle changes, virtual reality \\
E2 & Researcher (Psychology) & Personnel selection, clinical intervention, machine learning, natural language processing \\
E4 & Researcher (Psychology) & Digital mental health, smartphones, machine learning \\
E5 & Researcher (Psychology) & Cognitive neuroscience, physiology, eye-tracking \\
E6 & Inpatient Psychiatrist, Assistant Professor (Psychiatry) & Digital mental health, digital phenotyping, LLMs  \\
E7 & Researcher (Computer Science) & Digital health, ubiquitous computing, human-computer interaction \\
E8 & Researcher (Computer Science) & Digital mental health, ubiquitous computing, human-centered AI \\
E10 & Researcher (Psychology) & Passive smartphone sensing, machine learning, mood and behavior modeling \\
E12 & Researcher (Computer Science) & Stress detection, mobile health, machine learning, deep learning\\
E13 & Researcher (Computer Science) & Mobile sensing, digital health, eating disorders, self-control, emotion regulation \\
E14 & Researcher (Psychology) & Ecological momentary assessment, wearable sensing, eating disorders \\
E15 & Professor (Psychology) & Clinical child and adolescent psychology, developmental psychopathology, cognitive behavior therapy, social cognitive learning theory \\
E16 & Researcher (Psychology) & Healthy aging, daily experiences, multimorbidity \\
E17 & Assistant Professor (Psychology) & Emotion regulation, executive function, adolescent psychology \\
E18 & Assistant Professor (Psychology) & Ecological momentary assessment, intimate partner violence, substance use, dating violence, mindfulness \\
\hline
\end{tabular}
\end{table}

\section{List of Questions}
\subsection{General Questions:}
\begin{itemize}
    \item [1.] What traditional tools and techniques do you recommend to manage daily stress?
    \item [2.] What limitations do you see in these tools and techniques?
    \item [3.] What has been your experience with wearables and/or AI tools in stress or mental health contexts?
\end{itemize}

\subsection{Envisioning Integration}
\begin{itemize}
    \item[1.] What opportunities and risks do you see in integrating wearable sensing with LLM-based conversational systems in daily stress or general mental health contexts?
\end{itemize}

\subsection*{\normalfont\textit{EmBot DEMO}}

\subsection{Probe Interaction}
\begin{itemize}
    \item [1.] What are your initial impressions of this mobile app?
    \item [2.] How effectively do you think the mobile app aligns with your expectations for an integrated solution?
    \item [3.] What features of the app did you find most promising?
    \item [4.] What potential challenges, concerns, or risks stand out?
    \item [5.] What enhancements would you recommend for this mobile app?
\end{itemize}